# Scaling the propulsive performance of heaving and pitching foils

**Daniel Floryan,**† **Tyler Van Buren,**† **Clarence W. Rowley**† and **Alexander J. Smits**†

Scaling laws for the propulsive performance of rigid foils undergoing oscillatory heaving and pitching motions are presented. Water tunnel experiments on a nominally two-dimensional flow validate the scaling laws, with the scaled data for thrust, power, and efficiency all showing excellent collapse. The analysis indicates that the behaviour of the foils depends on both Strouhal number and reduced frequency, but for motions where the viscous drag is small the thrust closely follows a linear dependence on reduced frequency. The scaling laws are also shown to be consistent with biological data on swimming aquatic animals.

## 1. Introduction

The flow around moving foils serves as an abstraction of many interesting swimming and flight problems observed in nature. Our principal interest here is in exploiting the motion of foils for the purpose of propulsion, and so we focus on the thrust they produce, and their efficiency.

Analytical treatments of pitching and heaving (sometimes called plunging) foils date back to the early-mid 20[th] century. In particular, Garrick (1936) used the linear, inviscid, and unsteady theory of Theodorsen (1935) to provide expressions for the mean thrust produced by an oscillating foil, and the mean power input and output. Lighthill (1970) extended the theory to undulatory motion in what is called elongated-body theory. More recently, data-driven reduced-order modelling by, for example, Brunton *et al.* (2013), Brunton *et al.* (2014), and Dawson *et al.* (2015), has extended the range of validity and accuracy of similar models. A drawback of these treatments, however, is that it is often difficult to extract physical insights from them in regards to mean propulsive parameters such as thrust and efficiency.

In this respect, scaling laws can often prove valuable (Triantafyllou *et al.* 2005). In a particularly influential paper, Triantafyllou *et al.* (1993) established the importance of the Strouhal number in describing fishlike swimming flows, calling it the "dominant new parameter for fish propulsion" and "the governing parameter of the overall problem." The Strouhal number has since been adopted in nearly all subsequent works as the main parameter of interest (see, for instance, Quinn *et al.* (2014)), although the reduced frequency is sometimes preferred for foils with significant flexibility (Dewey *et al.* 2013).

Nevertheless, in conducting extensive experiments on pitching and heaving foils we find that such flows cannot be adequately described using only the Strouhal number or the reduced frequency. For instance, figure 1 shows the time-averaged thrust coefficient as a function of Strouhal number for a heaving foil. We see that the ratio of the heave amplitude to chord, $h^* = h_0/c$, has a significant impact on the thrust generated at a fixed Strouhal number. Here we report these findings, together with a new scaling analysis that helps to explain the experimental propulsive performance of rigid foils undergoing either heaving or pitching motions.

† Mechanical and Aerospace Engineering, Princeton University, Princeton NJ 08544, USA

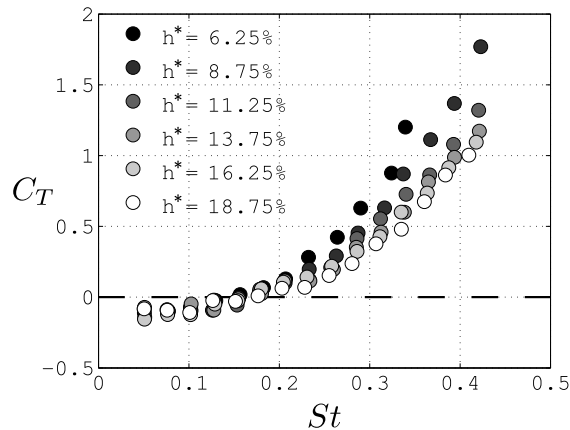

Figure 1: Time-averaged thrust coefficient $C_T$ as a function of Strouhal number $St$ for a heaving foil, for various heave amplitude to chord ratios, $h^*$. Experimental results from the current study. The parameters $C_T$ and $St$ are defined in Section 2.

## 2. Scaling laws

Consider a rigid two-dimensional foil moving at a constant speed $U_\infty$ while heaving and pitching about its leading edge. These motions are described by $h(t) = h_0 \sin(2\pi f t)$ and $\theta(t) = \theta_0 \sin(2\pi f t)$, respectively, where $h_0$ is the heave amplitude, $\theta_0$ is the pitch amplitude, and $f$ is the frequency; see figure 2. We are chiefly concerned with the time-averaged thrust in the streamwise direction produced by the foil motion, $F_x$, and the corresponding Froude efficiency

$$\eta = \frac{F_x U_\infty}{F_y \dot{h} + M \dot{\theta}}, \tag{2.1}$$

where $F_y$ is the force perpendicular to the freestream and $M$ is the moment taken about the leading edge. This efficiency is the ratio of power output to the fluid to power input to the foil. The relevant dimensionless parameters are

$$St = \frac{2fA}{U_\infty}, \qquad f^* = \frac{fc}{U_\infty}, \qquad A^* = \frac{A}{c},$$

where $St$ is the Strouhal number, $f^*$ is the reduced frequency, $A$ is the trailing edge amplitude of the motion ($h_0$ for heave, $c \sin \theta_0$ for pitch), and $c$ is the chord length of the foil. Although $St = 2f^* A^*$, we use all three parameters as a matter of convenience. Force and power coefficients are defined by

$$C_T = \frac{F_x}{\frac{1}{2}\rho U_\infty^2 sc}, \qquad C_y = \frac{F_y}{\frac{1}{2}\rho U_\infty^2 sc}, \qquad C_P = \frac{F_y \dot{h} + M \dot{\theta}}{\frac{1}{2}\rho U_\infty^3 sc},$$

where $\rho$ is the density of the surrounding fluid, and $s$ is the span of the foil.

We start with the notion that the forces acting on the foil are due to lift-based mechanisms, added mass effects, and viscous drag. We will assume that over our range of Reynolds numbers the drag coefficient $C_D$ is constant, independent of the amplitude or frequency of the actuation of the foil. This assumption will be justified by reference to the experimental data.

### 2.1. *Lift-based forces*

The only lift-based forces we consider are those that arise when the foil is at an instantaneous angle of attack to the freestream given by $\alpha = \theta - \arctan(\dot{h}/U_\infty)$. The effective flow velocity seen by the foil has a magnitude $U_{\text{eff}} = \sqrt{U_\infty^2 + \dot{h}^2}$, and an angle relative to the freestream velocity of $\arctan(\dot{h}/U_\infty)$. Hence,

$$F_x = -L\sin(\theta - \alpha) = -L\dot{h}/U_{\text{eff}},$$
$$F_y = L\cos(\theta - \alpha) = LU_\infty/U_{\text{eff}},$$

where $L$ is the lift on the foil given by $L = \frac{1}{2}\rho U_{\text{eff}}^2 sc C_L$, and the lift coefficient $C_L = 2\pi\sin\alpha + \frac{3}{2}\pi\dot{\alpha}c/U_\infty$ (Theodorsen 1935). The moment about the leading edge is $M = -cL/4$. Note that for a purely pitching foil, quasi-steady lift forces do not produce any thrust. High-frequency and large-amplitude motions will strengthen the nonlinearities in the response; the work of Liu *et al.* (2014) suggests that this will alter the phase differences between forces and motions. As such, terms that are expected to be $90°$ out of phase (for example, displacement and velocity, or velocity and acceleration) may develop in-phase components. These phase shifts are assumed to be constant for simplicity.

For heaving motions, neglecting viscous drag, we find

$$
\begin{aligned}
C_T &\sim 2\pi^3 St^2 + 3\pi^4 St^2 f^* U^*,\\
C_y &\sim 2\pi^2 St + 3\pi^3 St f^* U^*,\\
C_P &\sim 2\pi^3 St^2 + 3\pi^4 St^2 f^* U^*,\\
\eta &\sim 1,
\end{aligned}
\tag{2.2}
$$

where $U^* = U_\infty/U_{\text{eff}} = 1/\sqrt{1 + \pi^2 St^2}$. Similarly, for small pitching motions,

$$
\begin{aligned}
C_T &\sim 0,\\
C_y &\sim 2\pi A^* + \frac{3}{2}\pi^2 St,\\
C_P &\sim \frac{1}{2}\pi^2 St A^* + \frac{3}{8}\pi^3 St^2,\\
\eta &\sim 0.
\end{aligned}
\tag{2.3}
$$

In these scaling relations the first term is due to the angle of attack, and the second is due to the rate of change of the angle of attack. The $\sim$ sign indicates a proportionality, and although we expect the relative magnitudes of the first and second terms to be given by the analysis, the absolute magnitudes will need to be found by experiment.

### 2.2. *Added mass forces*

From Sedov (1965), the added mass forces per unit span on a flat plate are

$$
\begin{aligned}
F_t/s &= \rho\pi c'^2 V\dot{\theta} - \rho\pi c'^3 \dot{\theta}^2,\\
F_n/s &= -\rho\pi c'^2 \dot{V} + \rho\pi c'^3 \ddot{\theta},\\
M/s &= \rho\pi c'^3 \dot{V} - \frac{9}{8}\rho\pi c'^4 \ddot{\theta} - \rho\pi c'^2 UV + \rho\pi c'^3 U\dot{\theta},
\end{aligned}
$$

where $c' = c/2$, $U$ and $V$ are the instantaneous velocities in the directions tangential and normal to the plate, and subscripts $t$ and $n$ denote the instantaneous forces in the same directions. In our laboratory reference frame

$$F_x/s = \rho\pi c'^2 \left( \dot{h}\dot{\theta}\cos\theta - U_\infty\dot{\theta}\sin\theta - c'\dot{\theta}^2 \right)\cos\theta$$
$$+ \rho\pi c'^2 \left( \ddot{h}\cos\theta - \dot{h}\dot{\theta}\sin\theta - U_\infty\dot{\theta}\cos\theta + c'\ddot{\theta} \right)\sin\theta,$$
$$F_y/s = \rho\pi c'^2 \left( -\ddot{h}\cos\theta + \dot{h}\dot{\theta}\sin\theta + U_\infty\dot{\theta}\cos\theta + c'\ddot{\theta} \right)\cos\theta$$
$$+ \rho\pi c'^2 \left( \dot{h}\dot{\theta}\cos\theta - U_\infty\dot{\theta}\sin\theta - c'\dot{\theta}^2 \right)\sin\theta,$$
$$M/s = \rho\pi c'^2 \Big[ c' \left( \ddot{h}\cos\theta - \dot{h}\dot{\theta}\sin\theta - U_\infty\dot{\theta}\cos\theta \right) - \tfrac{9}{8}c'^2\ddot{\theta}$$
$$- \left( U_\infty\cos\theta + \dot{h}\sin\theta \right)\left( \dot{h}\cos\theta - U_\infty\sin\theta \right) + c'\left( U_\infty\cos\theta + \dot{h}\sin\theta \right)\dot{\theta} \Big].$$

Note that for a purely heaving foil, added mass does not produce any thrust.

For heaving motions, neglecting viscous drag, we find

$$\begin{aligned} C_T &\sim 0, \\ C_y &\sim \pi^3 St\, f^*, \\ C_P &\sim \pi^4 St^2\, f^*, \\ \eta &\sim 0. \end{aligned} \tag{2.4}$$

Similarly, for small pitching motions,

$$\begin{aligned} C_T &\sim \tfrac{1}{2}\pi^3 St^2 + \pi^2 St\, A^*, \\ C_y &\sim \tfrac{1}{2}\pi^3 St\, f^* + \tfrac{1}{2}\pi^2 St, \\ C_P &\sim \tfrac{9}{32}\pi^4 St^2\, f^* + \tfrac{1}{2}\pi^2 St\, A^*, \\ \eta &\sim \frac{16 + 16\pi f^*}{8 + 9\pi^2 f^{*2}}. \end{aligned} \tag{2.5}$$

In these scaling relations, the first term is the absolute added mass term, while the second term is due to being in a rotating frame of reference.

### 2.3. *Summary*

Combining lift-based and added mass forces, for purely heaving motions we have

$$\begin{aligned} C_T &= c_1 St^2 + c_2 St^2\, f^* U^* - C_{Dh}, \\ C_P &= c_3 St^2 + c_4 St^2\, f^* + c_5 St^2\, f^* U^*, \\ \eta &= \frac{c_1 + c_2 f^* U^*}{c_3 + c_4 f^* + c_5 f^* U^*}, \end{aligned} \tag{2.6}$$

where we have included the drag force for heaving motions ($C_{Dh}$) in the thrust scaling. The expression for the efficiency given here neglects the drag force, and so it should be interpreted as an inviscid scaling result. We will show the effects of viscous drag on the efficiency later in the text.

For purely pitching motions

$$\begin{aligned} C_T &= c_6 St^2 + c_7 St\, A^* - C_{Dp}, \\ C_P &= c_8 St^2 + c_9 St^2\, f^*, \\ \eta &= \frac{1}{f^*}\frac{c_6 f^* + c_7/2}{c_9 f^* + c_8}, \end{aligned} \tag{2.7}$$

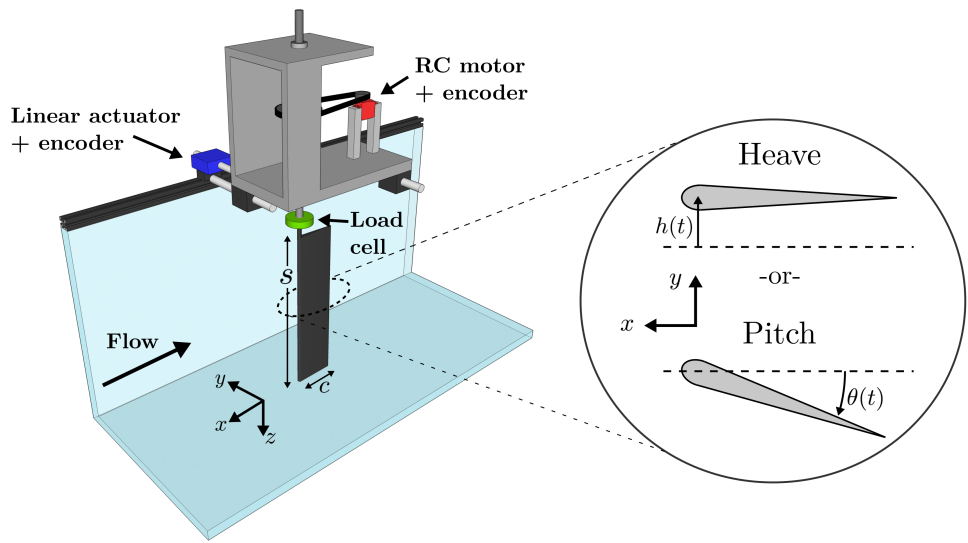

Figure 2: Experimental setup and sketch of motions.

where $C_{Dp}$ is the drag coefficient for pitching motions. The constants $c_1$ to $c_9$ will need to be found by experiment. Note that the expressions for efficiency only hold in the limit of negligible drag, so that they represent inviscid estimates.

## 3. Experimental setup

Experiments on a pitching or heaving foil were performed in a water tunnel. The foil was suspended in a free-surface recirculating water tunnel with a 0.46 m wide, 0.3 m deep, and 2.44 m long test section. The tunnel velocity was varied from 60 to 120 mm/s, with a typical turbulence intensity of 0.8%. A free surface plate was used to minimize the generation of surface waves. The experimental setup is shown in figure 2.

A teardrop foil was used for the experiments, with a chord of $c = 80$ mm, maximum thickness 8 mm, and span $s = 279$ mm, yielding an aspect ratio of $AR = 3.5$ and chord-based Reynolds number of $Re = 4780$ at 60 mm/s. To ensure that the flow was effectively two-dimensional, the gaps between the foil edges and the top and bottom surfaces of the water channel were less than 5 mm. Either pitching or heaving motions were used. An RC motor (Hitec HS-8370TH) was used to pitch the foil about its leading edge, and a linear actuator (Linmot PS01-23x80F-HP-R) was used to heave it on nearly frictionless air bearings (NewWay S301901). The pitch amplitude was varied from $\theta_0 = 3°$ to $15°$ in intervals of $2°$, the heave amplitude was varied from $h_0 = 5$ mm to 15 mm in intervals of 2 mm, and the frequency of actuation $f$ was chosen so that the Strouhal number varied from $St = 0.05$ to 0.4 in intervals of 0.025 (while maintaining $f < 2$ Hz). Pitch and heave motions were sampled continuously via encoders.

The foil thrust and efficiency were measured using a six-component force and torque sensor (ATI Mini40), which has force and torque resolutions of $5 \times 10^{-3}$ N and $1.25 \times 10^{-4}$ N·m in the $x$- and $y$-directions respectively, and $10^{-3}$ N and $1.25 \times 10^{-4}$ N·m in the $z$-direction. The force and torque data were acquired at a sampling rate of 100 Hz. During each experimental trial, the motion ran for 30 total cycles: the first five cycles were warm-up cycles, the following 20 cycles were for data acquisition, and the last five cycles were cool-down cycles. Each trial was run at least 6 times to ensure the repeatability of the data. Altogether, data were acquired for more than 1000 individual experiments.

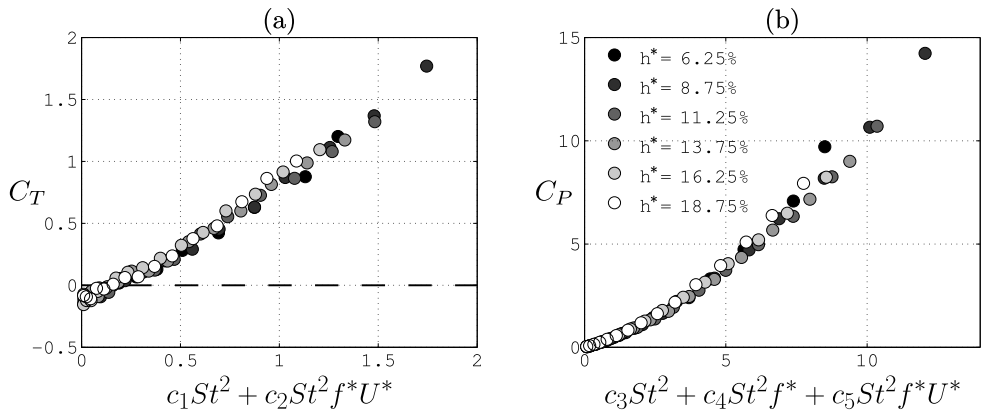

Figure 3: Heaving motions. Time-averaged (a) thrust and (b) power coefficients as functions of the scaling parameters (equation 2.6) for various $h^* = h_0/c$.

## 4. Heave results

Time-averaged thrust and power coefficients for the foil in heave are shown in figure 3. The data were taken at a fixed velocity of 60 mm/s. Performing a least-squares linear regression, the scaling constants were determined to be $c_1 = 3.52$, $c_2 = 3.69$, $c_3 = 27.47$, $c_4 = 13.81$, $c_5 = 5.06$, and a drag coefficient of 0.15 (equation 2.6). The values of the constants should not be taken to be universal; they are simply the values that work best for our data. The collapse of the data is relatively insensitive to the exact values of some of the constants, but the analysis indicates that each term in the model is of $O(1)$ importance, indicating that the physical mechanisms identified here are significant in explaining the data.

The thrust data collapse well onto a single curve, suggesting that the simplified physics used in our model are sufficient to explain the behaviour of the thrust. As shown in section 2, the $St^2$ term corresponds to the angle of attack, and the $St^2 f^* U^*$ term corresponds to the rate of change of angle of attack. We see therefore that the thrust for heaving motions is entirely due to lift-based forces, and that the effects of unsteadiness on the mean thrust are well captured by the rate of change of angle of attack.

Likewise, the power data collapse well onto a single curve, although there is some spread in the data for the stronger motions. The angle of attack, the rate of change of angle of attack, and added mass contribute to the power scaling. Power for heaving motions is thus affected by both lift-based and added mass forces, and the essential effects of unsteadiness on the mean power are well captured by the rate of change of angle of attack and added mass. It should be noted that the collapse of the mean power data is relatively insensitive to the values of the constants. The mean power is a weakly nonlinear function of the scaling parameter, suggesting the limits of our model; this is likely caused by the modification of the added mass (Liu *et al.* 2014).

The efficiency data are given in figure 4, presented as a function of Strouhal number (left), and as a function of the reduced frequency (right). For heaving motions, the scaling arguments indicate that the efficiency in the absence of drag should be approximately constant (for our constants and range of parameters). For higher values of the reduced frequency we observe that the efficiency data approach a constant, marked by a dashed line. The efficiency deviates from this trend for lower values of the reduced frequency and for smaller heave amplitudes due to the viscous drag on the foil. As motions become weaker, they produce less thrust. The drag, however, remains essentially constant. Thus as the motions become weaker, the drag will constitute a larger portion of the net

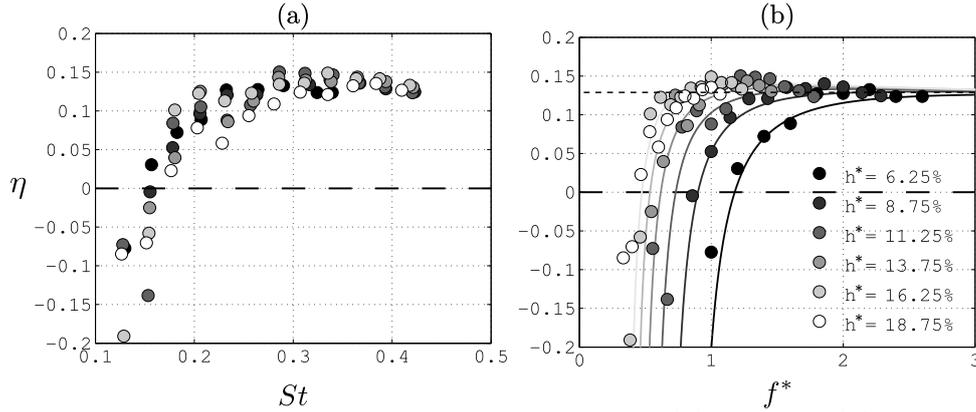

Figure 4: Heaving motions. Efficiency as a function of (a) $St$, and (b) and $f^*$. Solid lines indicate the scaling given by equation 2.6; dashed line indicates the scaling with $C_{Dh} = 0$.

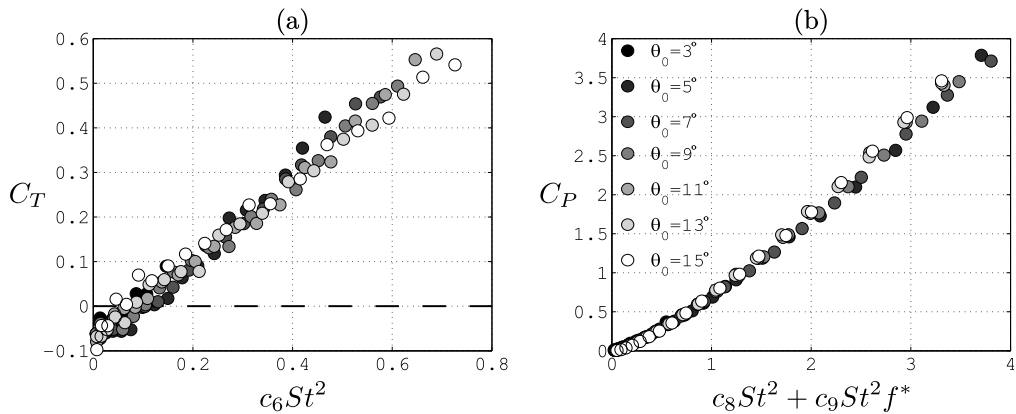

Figure 5: Pitching motions. Time-averaged (a) thrust and (b) power coefficients as functions of the scaling parameters (equation 2.7), with $c_7 = 0$.

streamwise force, eventually overtaking any thrust produced and leading to a negative efficiency.

## 5. Pitch results

Time-averaged thrust and power coefficients for pitching foils are shown in figure 5. The data were taken at a fixed velocity of 60 mm/s. Performing a least-squares linear regression, the scaling constants were determined to be $c_6 = 2.55$, $c_7 = 0$, $c_8 = 7.78$, $c_9 = 4.89$, and a drag coefficient of 0.08 (equation 2.7). Note that the thrust is affected by only the Strouhal number, as seen in previous work (Koochesfahani 1989). The values of the constants should be interpreted the same was as noted at the beginning of Section 4. The term multiplied by $c_7$ is negligible, indicating that it expresses the product of two terms (in this case displacement and velocity) that are 90° out of phase.

The thrust data follow our scaling model well. Although the pitch data are a bit more scattered than the heave data, the collapse is still evident, and the thrust coefficient varies linearly with the scaling parameter as expected. As shown in section 2, the $St^2$ term corresponds to added mass, and so the thrust for pitching motions is entirely due to added mass forces, which capture all of the effects of unsteadiness.

The power data also follow our scaling model well. Power in pitch is governed by the rate of change of angle of attack (the $St^2$ term), and by added mass forces (the $St^2 f^*$ term). These two terms alone capture the essential effects of unsteadiness on the mean power. As found for heaving motions, the mean power for pitch is a weakly nonlinear function of the scaling parameter, indicating the limits of our model.

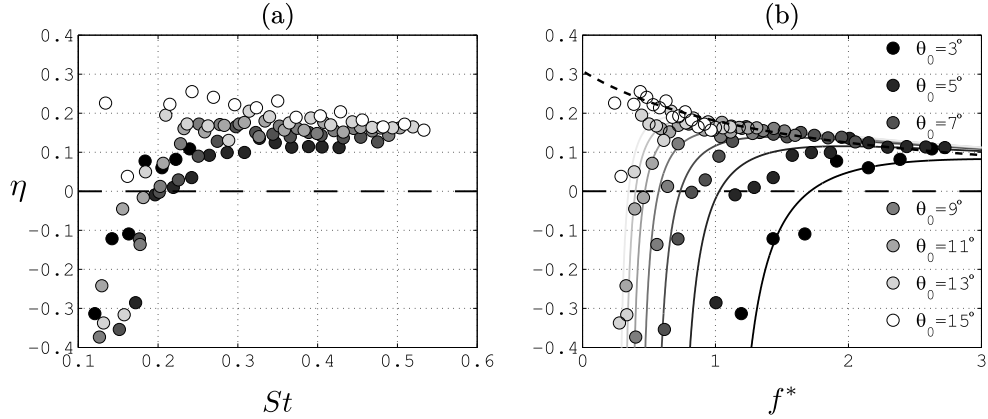

Figure 6: Pitching motions. Efficiency as a function of (a) $St$, and (b) $f^*$. Solid lines indicate the scaling given by equation 2.7; dashed line indicates the scaling with $C_{Dp} = 0$.

In figure 6 we present the efficiency data as a function of Strouhal number (left), and as a function of the reduced frequency (right). The scaling result in the absence of drag is given by $c_6/(c_8 + c_9 f^*)$ (see equation 2.7), shown in the figure by the dashed line. Clearly, the reduced frequency collapses the data for faster motions, whereas the Strouhal number does not. As found for heaving motions, the pitching data deviate from the scaling for slower motions where viscous drag becomes a significant portion of the net streamwise force. The solid lines show the curve fits after taking the drag into account. The inviscid scaling suggests that in order to maximize efficiency, the reduced frequency should be minimized, but viscous drag begins to be important at some point, and so for maximum efficiency an intermediate reduced frequency is best.

The inviscid scaling for the efficiency can be rewritten as $(c_6/c_9)/(c_8/c_9 + f^*)$. The efficiency curve thus behaves as $f^{*-1}$, but translated to the left by an amount $c_8/c_9$. The amount of leftward translation thus depends on the relative strengths of the terms corresponding to the coefficients $c_8$ (rate of change of angle of attack) and $c_9$ (added mass). From the perspective of maximizing efficiency, a smaller translation is better (e.g. $f^{*-1}$ without any translation approaches infinity as $f^*$ approaches zero). It is clear that we may alter the amount of translation, and thus the efficiency, by changing the relative strengths of lift-based and added mass forces. This could be achieved, possibly, by adding higher harmonics to the motion. This approach is currently under investigation.

## 6. Re-scaling thrust

If we consider motions where the viscous drag term is small, the thrust coefficients for heaving and pitching motions in equations 2.6 and 2.7 reduce to

$$C_{Th} = c_1 St^2 + c_2 St^2 f^* U^*,$$
$$C_{Tp} = c_6 St^2,$$

respectively, where we have taken $c_7 = 0$ as shown by the data. It is apparent that we may eliminate $St$ from both expressions by re-scaling the thrust, reducing the number of necessary scaling parameters. We rewrite the thrust laws as

$$C_{Th}^* = 4c_1 + 4c_2 f^* U^*, \tag{6.1}$$
$$C_{Tp}^* = 4c_6, \tag{6.2}$$

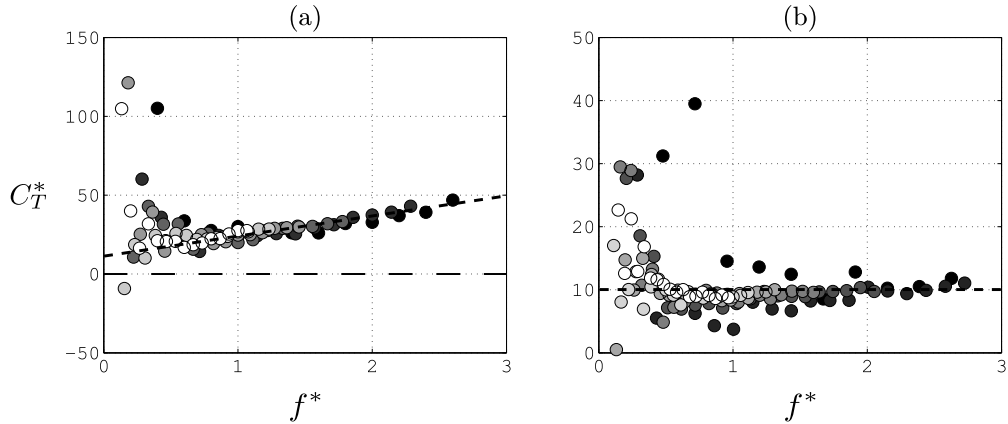

Figure 7: Newly non-dimensionalized thrust as a function of reduced frequency for (a) heaving and (b) pitching. Equations 6.1 and 6.2 are shown by the dashed lines. Colours are the same as in figures 3 and 5.

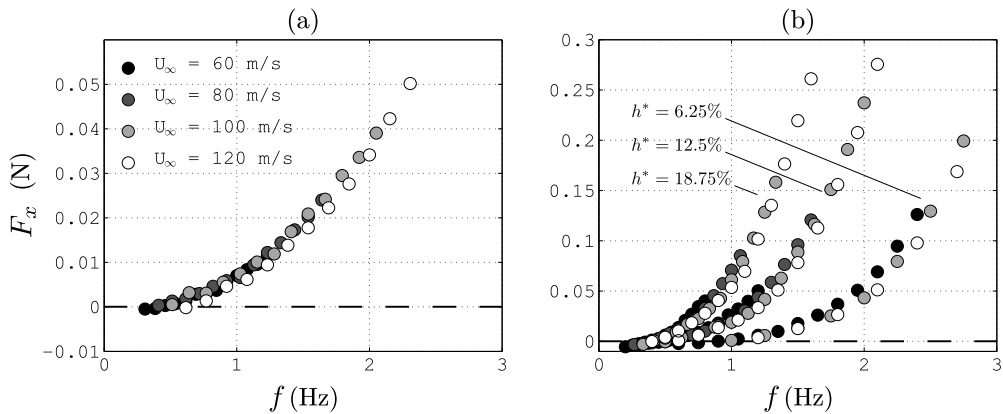

Figure 8: Dimensional thrust as a function of frequency at various freestream velocities for (a) pitching at $\theta_0 = 7°$ and (b) heaving at various $h^* = h_0/c$.

where we define a new thrust coefficient

$$C_T^* = \frac{F_x}{\frac{1}{2}\rho f^2 A^2 sc}$$

based on a characteristic velocity scale $fA$. Since $C_T^*$ does not contain the freestream velocity $U_\infty$, equation 6.2 indicates that for pitching motions the dimensional thrust should be independent of the freestream velocity.

The results, scaled as suggested by equations 6.1 and 6.2, are shown in figure 7. They show that the non-dimensional thrust coefficient $C_T^*$ is indeed a linear function of reduced frequency for heaving motions for large values of $f^*$ ($U^*$ varies only about 10% for our data, effectively constant), and a constant for pitching motions.

The experimental results presented thus far have all been taken at a single freestream velocity. To verify that the thrust results are truly independent of velocity, the velocity was varied from 60 to 120 mm/s. Figure 8 shows that the dimensional thrust is independent of freestream velocity for pitching motions, and only weakly depends on freestream velocity for heaving motions, confirming our scaling arguments. Although not shown here, we note that the data taken at different velocities follow the scaling laws given by equations 2.6 and 2.7, with the same values for the coefficients as found in Sections 4 and 5.

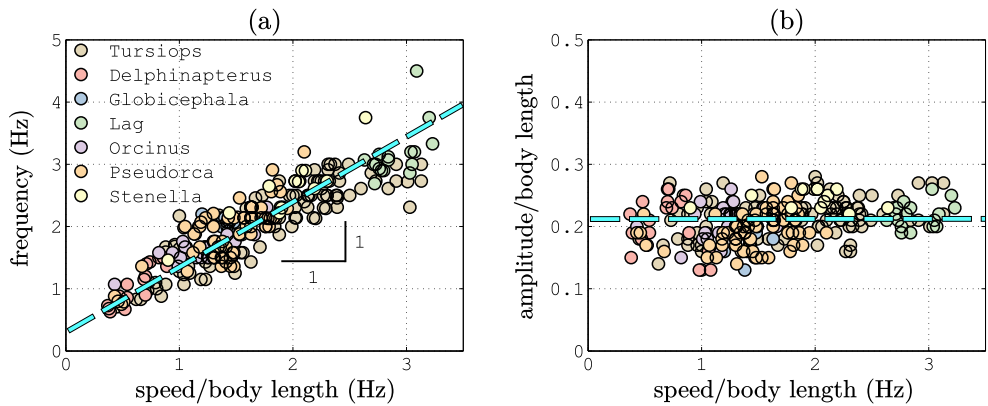

Figure 9: (a) Fluke-beat frequency and (b) non-dimensional fluke-beat amplitude as functions of length-specific swimming speed for several odontocete cetaceans. Adapted from Rohr & Fish (2004).

### 6.1. *Biological data*

It is instructive to test our scaling arguments against biological observations. Figure 9 shows fluke-beat frequency and non-dimensional fluke-beat amplitude as functions of length-specific swimming speed for several odontocete cetaceans (Rohr & Fish 2004). The data indicate that in order to increase their swimming speeds, these cetaceans increase their fluke-beat frequency while maintaining constant fluke-beat amplitude. In fact, their speeds increase at the same rate as their frequencies. In terms of non-dimensional variables, they are maintaining a constant reduced frequency.

Recall from section 2 that our scaling arguments indicate that the efficiency scales with $f^{*-1}$. Suppose that a swimmer wants to always swim as efficiently as possible. According to our scaling arguments, this corresponds to swimming at the value of reduced frequency which gives the greatest efficiency. Thus as a swimmer changes its speed, it must change its frequency accordingly in order to maintain the same reduced frequency. This is precisely what the biological data show.

## 7. Conclusions

Using only quasi-steady lift-based and added mass forces, new scaling laws for thrust coefficients, side force coefficients, power coefficients, and efficiencies were obtained for a rigid foil undergoing oscillatory heaving and pitching motions. The analysis indicates that the foil performance depends on both Strouhal number and reduced frequency. Water tunnel experiments on a nominally two-dimensional rigid foil showed that the scaling laws give an excellent collapse of the data. Viscous drag was seen to add an approximately constant negative offset to the thrust coefficient, but it causes the rapid decrease in efficiency seen for slower motions (low Strouhal number or small reduced frequency), and our scaling laws captured this behaviour well. For both heaving and pitching motions, the scaling indicates that slower motions lead to greater efficiency, as long as the motions are not so slow that viscous drag becomes a substantial component of the net streamwise force.

Biological observations of the swimming behaviour of odontocete cetaceans were shown to be consistent with our scaling arguments. When these aquatic creatures swim, they change their fluke-beat frequency in order to change their swimming speed while maintaining a constant fluke-beat amplitude. Under the premise of swimming as efficiently as possible, this behaviour of maintaining a constant reduced frequency is consistent with the scaling arguments presented.

Finally, observations of the weak dependence (or even independence) of dimensional thrust on freestream velocity led to the introduction of a new non-dimensionalization for thrust. The new non-dimensionalization reduces the thrust to only a linear function of reduced frequency for heaving motions, and to a constant value for pitching motions. The experimental data were shown to validate the new scaling, which is independent of Strouhal number.

This work was supported by ONR Grant N00014-14-1-0533 (Program Manager Robert Brizzolara). We would also like to thank Dr. Frank Fish for providing the biological data used in figure 9, and Dr. Keith Moored for many useful discussions.

## REFERENCES

BRUNTON, S. L., DAWSON, S. T. M. & ROWLEY, C. W. 2014 State-space model identification and feedback control of unsteady aerodynamic forces. *Journal of Fluids and Structures* **50**, 253–270.

BRUNTON, S. L., ROWLEY, C. W. & WILLIAMS, D. R. 2013 Reduced-order unsteady aerodynamic models at low reynolds numbers. *Journal of Fluid Mechanics* **724**, 203–233.

DAWSON, S. T. M., SCHIAVONE, N. K., ROWLEY, C. W. & WILLIAMS, D. R. 2015 A data-driven modeling framework for predicting forces and pressures on a rapidly pitching airfoil. In *45th AIAA Fluid Dynamics Conference*, pp. 1–14.

DEWEY, P. A., BOSCHITSCH, B. M., MOORED, K. W., STONE, H. A. & SMITS, A. J. 2013 Scaling laws for the thrust production of flexible pitching panels. *Journal of Fluid Mechanics* **732**, 29–46.

GARRICK, I. E. 1936 Propulsion of a flapping and oscillating airfoil. *Tech. Rep.* 567. NACA.

KOOCHESFAHANI, M. M. 1989 Vortical patterns in the wake of an oscillating foil. *AIAA Journal* **27** (9), 1200–1205.

LIGHTHILL, M. J. 1970 Aquatic animal propulsion of high hydromechanical efficiency. *Journal of Fluid Mechanics* **44** (02), 265–301.

LIU, T., WANG, S., ZHANG, X. & HE, G. 2014 Unsteady thin-airfoil theory revisited: Application of a simple lift formula. *AIAA Journal* **53** (6), 1492–1502.

QUINN, D. B., LAUDER, G. V. & SMITS, A. J. 2014 Scaling the propulsive performance of heaving flexible panels. *Journal of Fluid Mechanics* **738**, 250–267.

ROHR, J. J. & FISH, F. E. 2004 Strouhal numbers and optimization of swimming by odontocete cetaceans. *Journal of Experimental Biology* **207** (10), 1633–1642.

SEDOV, L. I. 1965 *Two-dimensional problems in hydrodynamics and aerodynamics*. Interscience Publishers.

THEODORSEN, T. 1935 General theory of aerodynamic instability and the mechanism of flutter. *Tech. Rep.* 496; originally published as ARR-1935. NACA.

TRIANTAFYLLOU, G. S., TRIANTAFYLLOU, M. S. & GROSENBAUGH, M. A. 1993 Optimal thrust development in oscillating foils with application to fish propulsion. *Journal of Fluids and Structures* **7** (2), 205–224.

TRIANTAFYLLOU, M. S., HOVER, F. S., TECHET, A. H. & YUE, D. K. 2005 Review of hydrodynamic scaling laws in aquatic locomotion and fishlike swimming. *Applied Mechanics Reviews* **58** (4), 226–237.